\documentclass[11pt,twoside,english,multicol,graphics]{article}

\usepackage{epsfig}

\usepackage[spanish]{babel}
\usepackage[latin1]{inputenc}
\usepackage{epsfig}
\begin{document}

\title{From time series to complex networks: the visibility graph\\Published in PNAS, vol. 105, no. 13 (2008) 4972-4975.}

 \author{Lucas Lacasa$^1$, Bartolo Luque$^1$, Fernando Ballesteros$^2$,\\ Jordi Luque$^3$, Juan Carlos Nuno$^4$\\
(1) Dpto. Matem\'{a}tica Aplicada y Estad\'{i}stica, ETSI
Aeron\'{a}uticos, \\Universidad Polit\'{e}cnica de Madrid, Spain\\
(2) Observatorio Astron\'{o}mico, Universidad de Valencia, Spain,\\
(3) Dept de Teoria del Senyal i Comunicacions, Universitat
Polit$\grave{e}$cnica de Catalunya, Spain\\
(4)Dpto. Matem\'{a}tica Aplicada a los Recursos Naturales, ETSI
Montes,\\ Universidad Polit\'{e}cnica de Madrid, Spain.}

\maketitle
\begin{abstract} In this work we present a simple and fast computational method, the
\emph{visibility algorithm}, that converts a time series into a
graph. The constructed graph inherits several properties of the
series in its structure. Thereby, periodic series convert into
regular graphs, and random series do so into random graphs.
Moreover, fractal series convert into scale-free networks, enhancing
the fact that power law degree distributions are related to
fractality, something highly discussed recently. Some remarkable
examples and analytical tools are outlined in order to test the
method's reliability. Many different measures, recently developed in
the complex network theory, could by means of this new approach
characterize time series from a new point of view. \end{abstract}




 In this letter we introduce a new tool in time series
analysis: the \emph{visibility graph}. This algorithm maps a time
series into a network. The main idea is to study into which extend
the techniques and focus of graph theory are useful as a way to
characterize time series. As will be shown below, this network
inherits several properties of the time series, and its study
reveals non trivial information about the series itself.

\begin{figure}[h]
 \hspace*{0.0cm}
 \includegraphics[width=0.70\textwidth]{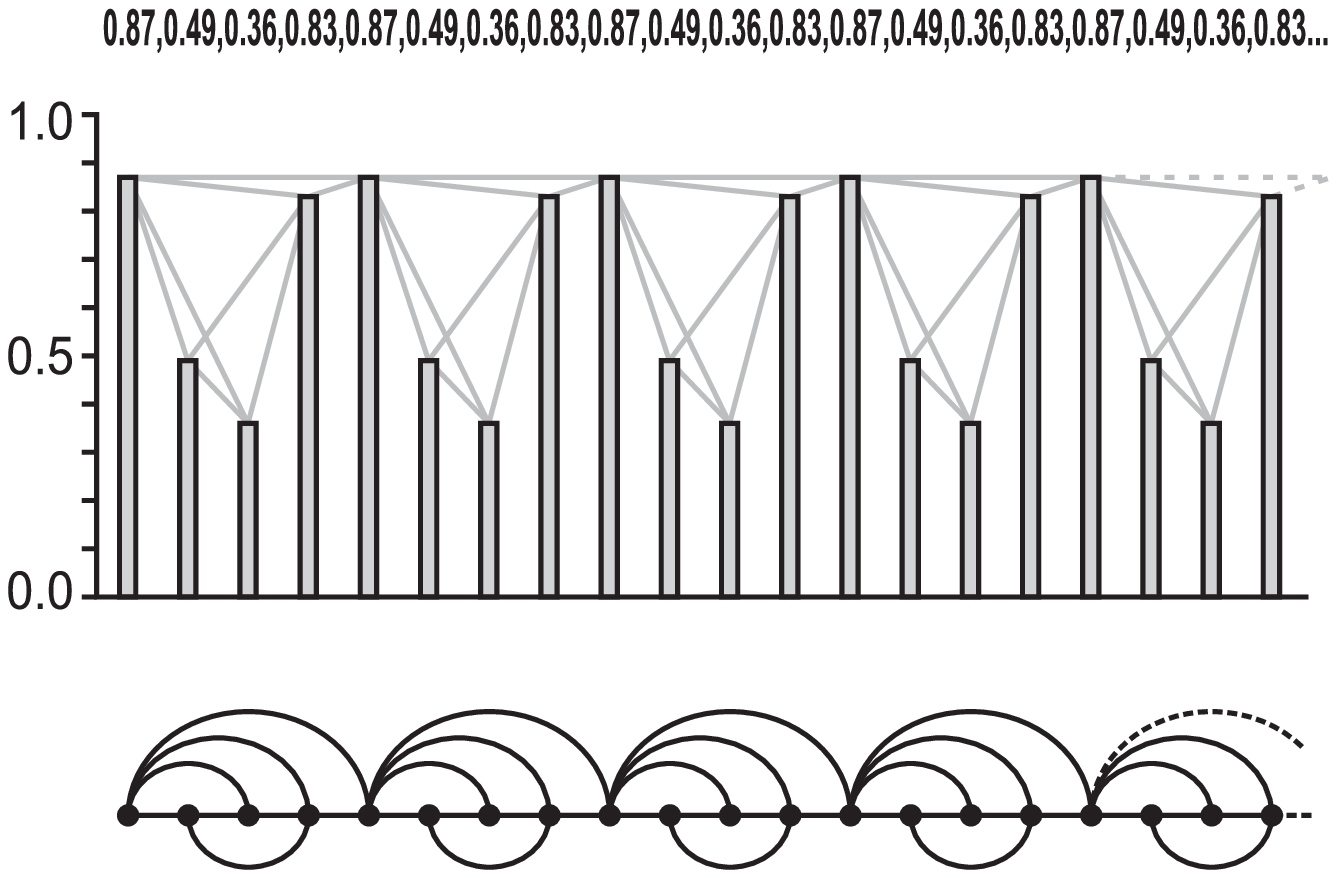}%
\caption{Example of a time series ($20$ data) and the associated
graph derived from the visibility algorithm. In the graph, every
node corresponds, in the same order, to a series data. The
visibility rays between the data define the links connecting nodes
in the graph.}
 \label{visibilidad_natural}
 \end{figure}

For illustrative purposes, in figure (\ref{visibilidad_natural}) we
present a scheme of the visibility algorithm. In the upper zone we
plot the first twenty values of a periodic series using vertical
bars (the data values are displayed above the plot). Considering
this as a landscape, we link every bar (every point of the time
series) with all those that can be seen from the top of the
considered one (gray lines), obtaining the associated graph (shown
in the lower part of the figure). In this graph, every node
corresponds, in the same order, to a series data, and two nodes are
connected if there exists visibility between the corresponding data,
that is to say, if there is a straight line that connects the series
data, provided that this ``visibility line'' does not intersect any
intermediate data height.

More formally, we can establish the following visibility criterium:
two arbitrary data $(t_a,y_a)$ and $(t_b,y_b)$ will have visibility,
and consequently will become two connected nodes of the associated
graph, if any other data $(t_c,y_c)$ placed between them fulfills:

\begin{equation}
y_c<y_b+(y_a-y_b)\frac{t_b-t_c}{t_b-t_a}. \label{crit}
\end{equation}

\begin{figure}[h]
 \hspace*{0.0cm}
 \includegraphics[width=0.70\textwidth]{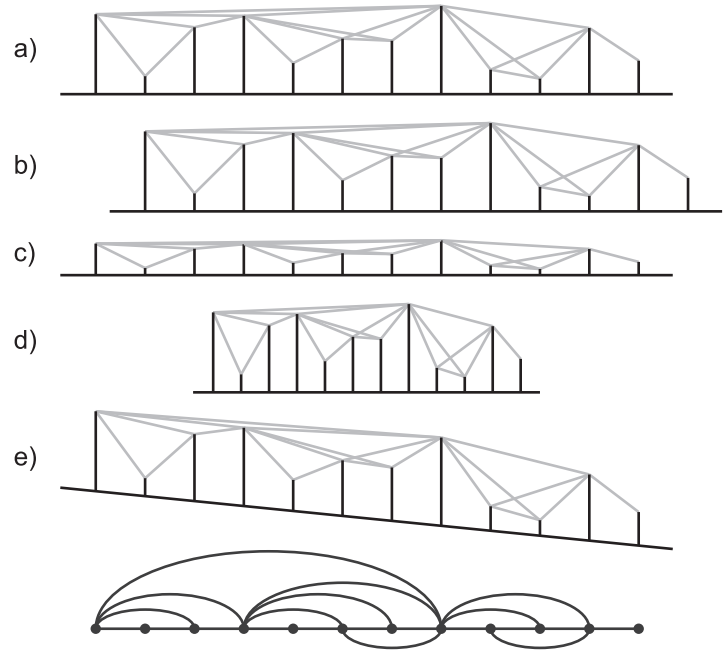}%
\caption{The visibility graph of a time series remains invariant
under several transformation of the time series: a) original time
series with visibility links b) translation of the data c) vertical
rescaling d) horizontal rescaling e) addition of a linear trend to
the data. As can be seen in the botom figure, in all these cases the
visibility graph remains invariant.}
 \label{invarianzas}
 \end{figure}

 We can easily check that by means of the present algorithm,
the associated graph extracted from a time series is always:\\
(\emph{i}) connected: each node sees at least its nearest
neighbors (left
and right).\\
(\emph{ii}) undirected: the way the algorithm is built up, there is
no direction defined in the links.\\
(\emph{iii}) invariant under affine transformations of the series
data: the visibility criterium is invariant under rescaling of both
horizontal and vertical axis, as well as under horizontal and
vertical translations (see figure \ref{invarianzas}).\\
In a recent work \cite{zhang}, Zhang $\&$ Small (ZS) introduced
another mapping between time series and complex networks. While the
philosophy is similar to this work (to encode the time series in a
graph in order to characterize the series using graph theory), there
exist fundamental differences between both methods, mainly in what
refers to the range of applicability (ZS only focus on
pseudoperiodic time series, associating each series cycle to a node
and defining links between nodes via temporal correlation measures,
while the visibility graph can be applied to every kind of time
series) and the graph connectedness (in ZS the giant component is
assured
only ad hoc, meanwhile the visibility graph is always connected by definition).\\

The key question is to know whether the associated graph inherits
some structure of the time series, and consequently if the process
which generated the time series may be characterized using graph
theory. In a first step we will consider periodic series. As a
matter of fact, the example plotted in figure
\ref{visibilidad_natural} is nothing but a periodic series with
period $4$. The associated visibility graph is regular, as long as
it is constructed by periodic repetition of a pattern. The degree
distribution of this graph is formed by a finite number of peaks
related to the series period, much in the vein of the Fourier Power
Spectrum of a time series. Generically speaking, all periodic time
series are mapped into regular graphs, the discrete degree
distribution being the fingerprint of the time series periods. In
the case of periodic time series, its regularity seems therefore to
be conserved or inherited structurally in the
graph by means of the visibility map.\\

As an opposite to periodic series, in a second step we will tackle a
series $R(t)$ of $10^6$ data extracted from an uniform distribution
in $[0,1]$. Although one would expect in a first moment a Poisson
degree distribution in this case (as for uncorrelated random graphs
\cite{random_graphs}), a random time series has indeed some
correlation, since it is an ordered set. In fact, let $k_t$ be the
connectivity of the node associated to the data $t$. If $k_t$ is
large (related to the fact that the data has a large value and that
consequently it has large visibility), one would expect that
$k_{t+1}$ would be relatively small, since the time series is random
and two consecutive data with a large value are not likely to occur.
It is indeed due to these 'unlikely' large values (the hubs) that
the tail of the degree distribution deviates from the Poisson
distribution. Two large values in the series data can be understood
as two rare events in the random process. The time distribution of
these events is indeed exponential \cite{Feller, nota}, therefore we
should expect the tail of the degree distribution in this case to be
exponential instead of Poissonian, as long as the form of this tail
is related to the hub's distribution.

\begin{figure}[h]
 \hspace*{-0.5cm}
\includegraphics[width=0.70\textwidth]{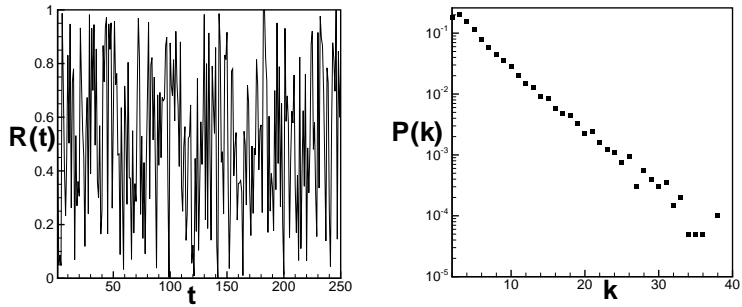}%
\caption{Left figure: First 250 values of $R(t)$, where $R$ is a
 random series of $10^6$ data extracted from U[0,1]. Right figure: degree
distribution $P(k)$ of the visibility graph associated to $R(t)$
(plotted in semi-log). While the beginning of the curve approaches
the result of a Poisson process, the tail is clearly exponential.
This behavior is due to data with large values (rare events), which
are the hubs.}
 \label{random}
 \end{figure}

In the left side of figure \ref{random} we depict the first 250
values of $R(t)$. In the right side we plot the degree distribution
$P(k)$ of its visibility graph. The tail of this distribution fits
quite well an exponential distribution, as expected. Note at this
point that time series extracted randomly from other distributions
than uniform have also been addressed. In every case the algorithm
captures the random nature of the series, and the particular shape
of the degree distribution of the visibility graph is related to the
particular random process \cite{nota}.

 Hitherto, ordered (periodic) series convert into regular graphs,
 and random series convert into exponential random graphs:
 order and disorder structure in the time series seem to be inherited in the topology of the
 visibility graph. Thus, the question arises: What kind of visibility graph is
obtained from a fractal time series? This question is in itself
interesting at the present time. Recently, the relationship between
self-similar and scale-free networks \cite{scalefree, reviewnet,
reviewnet2, reviewnet3, reviewnet4} has been intensively discussed
\cite{nature_havlin,PRL_skeleton,nature2_havlin,PRE_skeleton}.
 While complex networks \cite{reviewnet} usually exhibit the Small-World property \cite{SW} and
 cannot be consequently size-invariant, it has
been recently shown \cite{nature_havlin} that applying fitted
box-covering methods and renormalization procedures, some real
networks actually exhibit self-similarity. So, whereas
self-similarity seems to imply scale-freeness, the
opposite is not true in general.\\

In order to explore these issues in more detail, the following two
fractal series will be considered: the well-known Brownian motion
$B(t)$ and the Conway series \cite{CONWAY}. While the Brownian
motion represents a well-known case of self-affinity (indeed, the
following relation holds: $B(t)=a^{1/2}B(t/a)$), the Conway series
$a(n)-n/2$ is the recursively generated fractal series from:

\begin{eqnarray}
&&a(1)=a(2)=1 \nonumber \\
&&a(n)=a(a(n-1))+ a(n-a(n-1));\  n>2.\nonumber\\
\end{eqnarray}

\begin{figure}[h]
 \hspace*{-0.5cm}
 \includegraphics[width=0.70\textwidth]{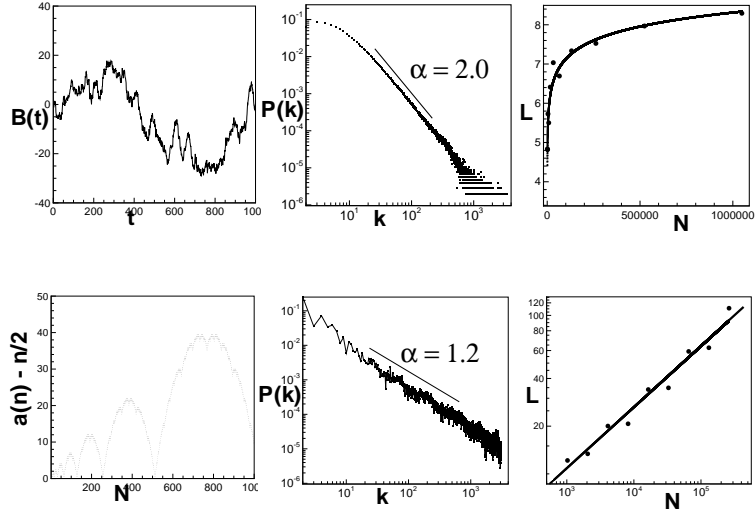}%
 \caption{Upper part, from left to right: First 4000 data from a Brownian series of
$10^6$ data. In the middle, the degree distribution of the
visibility graph associated
 to the Brownian motion. This one is a power law $P(k)\sim
k^{-\alpha}$ with $\alpha=2.00\pm0.01$. In the right part of the
figure we plot the mean path length of this network as a function of
the network size N. The best fitting provides a logarithmic scaling
$L(N)=1.21+0.51\log(N)$. This network shows Small-World effect in
addition to be scale-free. Bottom part, from left to right: First
$10^5$ data from a Conway series of $4\cdot10^6$ data. In the
middle, the degree distribution of the visibility graph associated
 to the Conway series. This one is a power law $P(k)\sim k^{-\alpha}$ with $\alpha=1.2\pm0.1$.
 The mean path length as a
function of the size N is depicted in the right part of the figure.
The best fitting provides a power law scaling $L(N)=0.76N^{0.38}$.
Then, this network is scale-invariant.} \label{colapsed}
\end{figure}

In figure \ref{colapsed} we have plotted the behavior of these
series, the degree distribution $P(k)$ of their respective
visibility graphs and their mean path length $L(N)$ as a function of
the series length. First, both series have visibility graphs with
degree distributions that correspond to power laws of the shape
$P(k)\sim k^{-\alpha}$, where we get different exponents in each
case: this result enhances the fact that in the context of the
visibility algorithm, power law degree distributions (that is, scale
free networks \cite{reviewnet, reviewnet2, reviewnet3, reviewnet4})
arise naturally from fractal series. Moreover, this relation seems
to be robust as long as the preceding examples show different kinds
of fractality: while $B(t)$ stands for a stochastic self-affine
fractal, the Conway series is a deterministic series recursively
generated. On the other hand, while the Brownian visibility graph
seems to evidence the Small-World effect (right top figure
\ref{colapsed}) as $L(N)\sim\log(N)$, the Conway series shows in
turn a self-similar relation (right bottom figure \ref{colapsed}) of
the shape $L(N)\sim N^\beta$. This fact can be explained in terms of
the so called hub repulsion phenomenon \cite{nature2_havlin}:
visibility graphs associated to stochastic fractals such as the
Brownian motion or generic noise series do not evidence repulsion
between hubs (in these series it is straightforward that the data
with highest values would stand for the hubs, and these data would
have visibility between each other), and consequently won't be
fractal networks following Song et al. \cite{nature2_havlin}. On the
other hand, the Conway series actually evidence hub repulsion: this
series is concave-shaped and consequently the highest data won't in
any case stand for the hubs; the latter ones would be located much
likely in the monotonic regions of the series, which are indeed
hidden from
each other (effective repulsion) across the series local maxima. The Conway visibility graph is thus fractal.\\
Since a fractal series is characterized by its
Hurst exponent, we may argue that the visibility graph can actually
distinguish different types of fractality, something that will be
explored in detail in further work.
 Note at this point that some other fractal series have been also
studied (Q series \cite{QSERIES}, Stern series \cite{STERN},
Thue-Morse series \cite{Thue-Morse}, etc) with similar results.
Moreover, observe that if the series under study increases its
length, the resulting visibility graph can be interpreted in terms
of a model of network growth and may eventually shed light into the
fractal network formation problem.\\

\begin{figure}[h]
 \hspace*{-0.5cm}
 \includegraphics[width=0.70\textwidth]{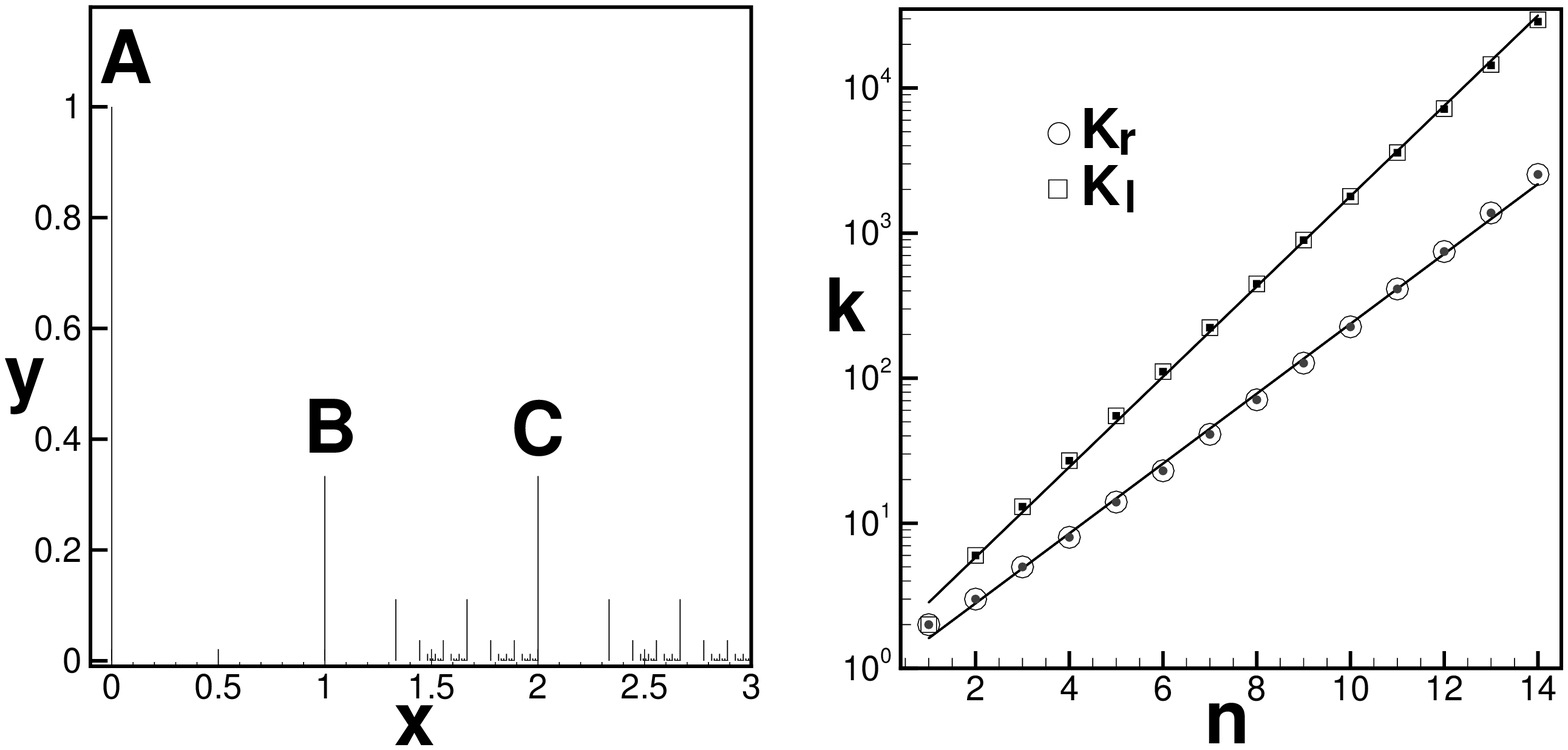}%
 \caption{Left: Fractal series obtained by iteration of the original pattern
 (points A,B,C) with $p=10$ steps. Right: Values of $K_r$ (circles) and $K_l$ (squares)
 as a function of the fractal size, related to
 equations (\ref{c-right},\ref{c-left}). Note that the plot is log-linear:
 the relation is thus exponential.
 The straight lines correspond to the approximations deduced in
 equations (\ref{c-right-approx},\ref{c-left-approx}).}
 \label{patron_AUTOSIM}
 \end{figure}

 In order to cast light into the relation between fractal series
and power law distributions, in the left part of figure
\ref{patron_AUTOSIM} we present a deterministic fractal series
generated by iteration of a simple pattern of three points. The
series starts (step 0) with three points (A,B,C) of coordinates
$(0,1)$, $(1,1/3)$ and $(2,1/3)$ respectively. In step \emph{p}, we
introduce $2^{p+1}$ new points with height $3^{-p-1}$ and distanced
$3^{-p}$. The series form a self-similar set: applying an isotropic
zoom of horizontal scale $3^p$ and vertical scale $3^p$ to the
pattern of order \emph{p}, we recover the original pattern.

 Note that this time series is not
 data uniformly spaced as the previous examples. However its usefulness is set on the fact that it is simple enough to
 handle it analytically, that is to
 say, to find the degree distribution of its visibility graph.
 The main idea is to find a recurrence behavior in the way that a given node increases
 its connectivity when the fractal step (that is to say, the fractal size) is increased \cite{BARABASI}.
 Then we calculate how many
 nodes (self-similar to it) appear in each step, and from both
 relations we come to a degree distribution for these kind of nodes.\\
 First, from a quick visual exploration of left figure \ref{patron_AUTOSIM} one comes to
 the conclusion that nodes A and B have typically the same degree.
 In the other hand, the degree of node C can be decomposed in two
 terms, the left degree (due to visibility of nodes at the left of
 C) and right degree. The degree of A and B is the same as the right
 degree of C (statistically speaking, A and B increase their connectivity as the fractal size increases much in the
 way of the right part of C). Thereby, the degree of C provides the whole
 information of the system. We will quote $K_r(C,n)$ the right degree of node C in a n-step fractal
 (respectively, $K_l(C,n)$ stands for the left degree).\\
 Applying the visibility criterium, one can geometrically find that
 \begin{equation}
K_r(C,n)=\sum_{m=1}^n\frac{1}{m}\sum_{d|m}\mu(d)\cdot2^{m/d},
 \label{c-right}
\end{equation}
where $\mu$ is the Mo\"{e}bius function. Note that this summation
agrees with the number of irreducible polynomials of degree at most
$n$ over the Galois field GF(2) \cite{galois}, something which
deserves an in-depth investigation. This expression can be
approximated by
\begin{equation}
K_r(C,n)\sim 2^{4n/5}.\label{c-right-approx}
\end{equation}
On the other hand, there is a recurrence in the left degree that
reads
\begin{eqnarray}
K_l(C,n)=2K_l(C,n-1) + 1, \label{c-left}
\end{eqnarray}
whose leading term is
\begin{equation}
K_l(C,n)\sim 2^n. \label{c-left-approx}
\end{equation}

The node $C$ will thus have a degree $K(C)=K_r(C,n)+K_l(C,n)$. In
figure \ref{patron_AUTOSIM} (right) we plot the values of $K_r$
(circles) and $K_l$ (squares) as a function of the fractal size (the
number of iterations $n$). Numerical values are plotted as the outer
circles and squares, while the inner circles and squares come from
plotting equations (\ref{c-right},\ref{c-left}). Note that both
formulas reproduce the numerical data. The straight lines correspond
to the approximation equations (\ref{c-right-approx}) and
(\ref{c-left-approx}). Now, in a generic step \emph{p}, $2^p$ nodes
which are self-similar to C appear (by construction). Those nodes
will have a degree $K(C,n-p)=2^{\frac{4}{5}(n-p)}+2^{n-p}$ that, for
large values of $n-p$, can be approximated to
$K(C,n-p)\simeq2^{n-p}$ . Defining $f(K)$ as the degree
distribution, we get that $f\big(K(C,n-p)\big)=2^p$,
 and with the change of variable $ u\equiv 2^{n-p}$, it is easy to
come into:
\begin{equation}
f(u)\sim u^{-1},
\end{equation}
that is, the degree distribution related to the C-nodes is a power law.\\
Although this simple example doesn't provide a general explanation
of why fractality is traduced into power law distributions, it may
stand as a generic way of dealing with deterministic fractal series
generated by iteration.\\

Once the visibility method has been presented, some remarks can be
stated: note that typically two series that only differ by an affine
transformation will have the same visibility graph; in this sense
the algorithm absorbs the affine transformation. Furthermore, it is
straightforward to see that that some information regarding the time
series is inevitably lost in the mapping from the fact that the
network structure is completely determined in the (binary) adjacency
matrix. For instance, two periodic series with the same period as
$T_1=\{..,3,1,3,1,..\}$ and $T_2=\{..,3,2,3,2,..\}$ would have the
same visibility graph albeit being quantitatively different. While
the spirit of the visibility graph is to focus on time series
structural properties (periodicity, fractality, etc), the method can
be trivially generalized using weighted networks (where the
adjacency matrix isn't binary and the weights determine the slope of
the visibility line between two data) if we eventually need to
quantitatively
distinguish time series like $T_1$ and $T_2$ for instance.\\

While in this paper we have only tackled undirected graphs, note
that one could also extract a directed graph (related to the
temporal axis direction) in such a way that for a given node one
should distinguish two different connectivities: an ingoing degree
$k_{in}$, related to how many nodes see a given node $i$, and an
outgoing degree $k_{out}$, that is the number nodes that node $i$
sees. In that situation, if the direct visibility graph extracted
from a given time series is not invariant under time reversion (that
is, if $P(k_{in})\neq P(k_{out})$), one could assert that the
process that generated the series is not conservative. In a first
approximation we have studied the undirected version and the
directed one will be
eventually addressed in further work.\\

There are some direct applications of the method that can be put
forward. The relation between the exponent of the degree
distributions and the Hurst exponent of the series will be addressed
in further work. In particular, it turns out that the method
presented here constitutes a reliable tool to estimate Hurst
exponents, as far as a functional relation between the Hurst
exponent of a fractal series and the degree distribution of its
visibility graph holds \cite{further}. Note that the estimation of
Hurst exponents is an issue of major importance in data analysis
that is yet to be completely solved (see for instance \cite{hurst}).
Fractional Brownian motions, a concept of great interest in a large
variety of fields ranging from electronic devices to Biology, will
also be considered
in relation with the preceding point.\\
Moreover, the ability of the algorithm to detect not only the
difference between random and chaotic series but also the spatial
location of inverse bifurcations in chaotic dynamical systems is
another fundamental issue that will also be at the core of further
investigations \cite{further}. Finally, the visibility graph
characterizes non trivial time series and in that sense, the method
may be relevant in specific problems of different garments, such as
human behavior time series recently
put forward \cite{human}.\\

In summary, a brand new algorithm that converts time series into
graphs is presented. The structure of the time series is conserved
in the graph topology: periodic series convert into regular graphs,
random series into random graphs and fractal series into scale-free
graphs. Such characterization goes beyond, since different graph
topologies arise from apparently similar fractal series. In fact,
the method captures the hub repulsion phenomenon associated to
fractal networks \cite{nature2_havlin} and thus distinguishes scale
free visibility graphs evidencing Small-World effect from those
showing scale invariance.\\
With the visibility algorithm a natural bridge between complex
networks theory and time series analysis is now built.\\

We want to acknowledge the comments from the editor and two
anonymous referees. The research was supported by grant
FIS2006-08607 from the Spanish Ministry of Science.



\newpage

\newpage

\newpage

\newpage

\newpage

\begin{thebibliography}{}
\bibitem{zhang} J. Zhang, M. Small, \textit{Phys. Rev. Lett.} \textbf{96}, 238701
(2006).
\bibitem{random_graphs} B. Bollob\'as, \textit{Modern Graph
Theory}, Springer-Verlag, New York Inc. (1998).

\bibitem{Feller} W. Feller, \textit{An Introduction to Probability Theory and its
Applications}, John Willey and Sons, Inc. (1971).

\bibitem{nota} This feature will be adressed in further work.


\bibitem{scalefree} A.L. Barab\'asi, R. Albert, \textit{Science}
\textbf{286}, 509 (1999).
\bibitem{nature_havlin} C. Song, S. Havlin, H.A. Makse, \textit{Nature} \textbf{433}, 392 (2005).
\bibitem{PRL_skeleton}K.I. Goh, G. Salvi, B. Kahng, D. Kim, \textit{Phys. Rev. Lett.} \textbf{96} (2006).
\bibitem{nature2_havlin} C. Song, S. Havlin, H.A. Makse, \textit{Nat. Phys.} \textbf{2},275 (2006).
\bibitem{PRE_skeleton}J.S.Kim, K.I. Goh, G. Salvi, E. Oh, B. Kahng, D. Kim, \textit{Phys. Rev. E} \textbf{75}, 016110 (2007).
\bibitem{reviewnet} R. Albert, A.L. Barab\'asi, \textit{Rev. Mod. Phys.} \textbf{74} (2002).
\bibitem{reviewnet2} M.E.J. Newman, \textit{SIAM Review} \textbf{45} (2003) 167-256.
\bibitem{reviewnet3} S. Dorogovtsev, J.F.F Mendes, \textit{Advances in Physics} \textbf{51}, 4 (2002).
\bibitem{reviewnet4} S. Bocaletti, V. Latora, Y. Moreno, M. Ch\'avez and D.U. Hwang, \textit{Phys. Reports} \textbf{424} (2006) 175-308.
\bibitem{SW} D.J. Watts and S.H. Strogatz, \textit{Nature} \textbf{393}, 440-442 (1998).




\bibitem{CONWAY}J. Conway, \textit{Some Crazy Sequences}, Lecture at AT$\&$T Bell Labs (1988).

\bibitem{QSERIES}D. Hofstadter, \textit{G\"{o}del, Escher, Bach},
New York: Vintage Books, pp. 137-138 (1980).

\bibitem{STERN}M.A. Stern, \textit{J. Reine Angew. Math.}, 55, pp. 193-220 (1858).
\bibitem{Thue-Morse}M.R. Schroeder, \textit{Fractals, Chaos, and Power Laws}, New York: W. H. Freeman (1991).

\bibitem{BARABASI} A.L. Barab\'asi, E. Ravasz, and T. Vicsek, \textit{Physica A} 299, 559-564
(2001).
\bibitem{galois} K.H. Hicks, G.L. Mullen, and I. Sato, \textit{Distribution of irreducible polynomials
over $F_2$}, in Finite Fields with Applications to Coding Theory,
Cryptography and Related Areas (Oaxaca, 2001), 177-186, Springer,
Berlin, 2002.
\bibitem{further} To be published.
\bibitem{hurst} T. Karagiannis, M. Molle and M. Faloutsos, \textit{Internet
Computing, IEEE} \textbf{8}, 5 (2004).
\bibitem{human} A. V\'azquez, J. Gama Oliveira, Z. Desz$\ddot{o}$, K. Goh, I. Kondor and A.L. Barab\'asi, \textit{Phys. Rev. E} \textbf{73}, 036127 (2006).

\end{thebibliography}
\end{document}